\pdfoutput=1

\documentclass[11pt,twoside,a4paper,cmspaper,final,collab]{cms-tdr}
\def\svnVersion{246096}\def\cmsCernNo{2013-200}\def\cmsCernDate{\today}\def\cmsMessage{Published in the European Physical Journal C as \href{http://dx.doi.org/10.1140/epjc/s10052-014-2901-8}{\doi{10.1140/epjc/s10052-014-2901-8.}}}

\begin{document}\cmsNoteHeader{SMP-12-010}

\hyphenation{had-ron-i-za-tion}
\hyphenation{cal-or-i-me-ter}
\hyphenation{de-vices}
\RCS$Revision: 237279 $
\RCS$HeadURL: svn+ssh://svn.cern.ch/reps/tdr2/papers/SMP-12-010/trunk/SMP-12-010.tex $
\RCS$Id: SMP-12-010.tex 237279 2014-04-18 22:43:37Z alverson $
\ifthenelse{\boolean{cms@external}}{\providecommand{\breakhere}{\linebreak[4]}}{\providecommand{\breakhere}}{\relax}
\newlength\cmsFigWidth
\ifthenelse{\boolean{cms@external}}{\setlength\cmsFigWidth{\columnwidth}}{\setlength\cmsFigWidth{0.9\textwidth}}
\ifthenelse{\boolean{cms@external}}{\providecommand{\cmsLeft}{top}}{\providecommand{\cmsLeft}{left}}
\ifthenelse{\boolean{cms@external}}{\providecommand{\cmsRight}{bottom}}{\providecommand{\cmsRight}{right}}
\cmsNoteHeader{SMP-12-010} 
\title{Probing color coherence effects in pp collisions at $\sqrt{s} = 7\TeV$}

\date{\today}
\abstract{
A study of color coherence effects in $\Pp\Pp$ collisions at a center-of-mass energy of 7\TeV is presented.
The data used in the analysis were collected in 2010 with the CMS detector at the LHC and correspond to an integrated luminosity of $36\pbinv$.
Events are selected that contain at least three jets and where the two jets with the largest transverse momentum exhibit a back-to-back topology.
The measured angular correlation between the second- and third-leading
jet is shown to be sensitive to color coherence effects, and is
compared to the predictions of Monte Carlo models with various
implementations of color coherence.
None of the models describe the data satisfactorily.
}
\hypersetup{%
pdfauthor={CMS Collaboration},%
pdftitle={Probing color coherence effects in pp collisions at sqrt(s) = 7 TeV},%
pdfsubject={CMS},%
pdfkeywords={CMS, physics, QCD, jet, color coherence}}
\ifthenelse{\boolean{cms@external}}{
\titlerunning{Probing color coherence effects in pp collisions at $\sqrt{s} = 7\TeV$}
}{}
\maketitle 
\section{Introduction} \label{sec-intro}
An important feature of the color interaction in quantum chromodynamics (QCD) is that the outgoing partons produced in the hard interaction continue to interfere with each other during their fragmentation phase.
This phenomenon, called \textit{color coherence}, manifests itself by the relative abundance of soft radiation in the region between the color connected final-state partons and the suppression of soft radiation elsewhere.

Color coherence phenomena were initially observed in \EE collisions by several experiments at PETRA, PEP and LEP~\cite{JADE, TASSO, Aihara:1984du, Aihara:1984mn, TPC, MARK2,OPAL,L3}.
These experiments showed the coherence effect in $\EE\rightarrow\cPq\cPaq\cPg$ three-jet events through the suppression of particle production in the region between the quark and antiquark jets.

In hadron collisions, in addition to the color connection between the final-state partons, the color connection between the outgoing partons and the incoming partons must be considered.
The Tevatron experiments CDF and \DZERO have both reported evidence for color coherence effects in measurements of the spatial correlations between neighboring jets~\cite{CDF,D0}.
These correlations were not well reproduced by Monte Carlo (MC) simulations that use incoherent parton shower models.
However, the data were successfully described by simulations that include color coherence effects through the ordering of the parton emission angles~\cite{PS-0}.

The technique originally developed by the Tevatron experiments is used to study color coherence effects in pp collisions at $\sqrt{s}=7$\TeV with the Compact Muon Solenoid (CMS) detector. Events with at least three jets (called three-jet events) are selected, and these jets are ordered by their transverse momenta $p_\mathrm{T1} > p_\mathrm{T2} > p_\mathrm{T3}$ with respect to the beam direction. We measure the angular correlation between the second and third jet to probe the effects of color coherence.

The CMS detector has a right-handed coordinate system with its origin at the center of the detector.
The $z$ axis points along the direction of the counterclockwise beam, $\phi$ is the azimuthal angle in
the transverse plane perpendicular to the beam, and $\theta$ is the polar angle relative to the $z$ axis.
The pseudorapidity of the $i$th jet is denoted by $\eta_i = - \ln [\tan(\theta_i/2)]$ and its azimuthal angle by $\phi_i$.

\begin{figure}[htb]
\begin{center}
{\includegraphics[width=\cmsFigWidth]{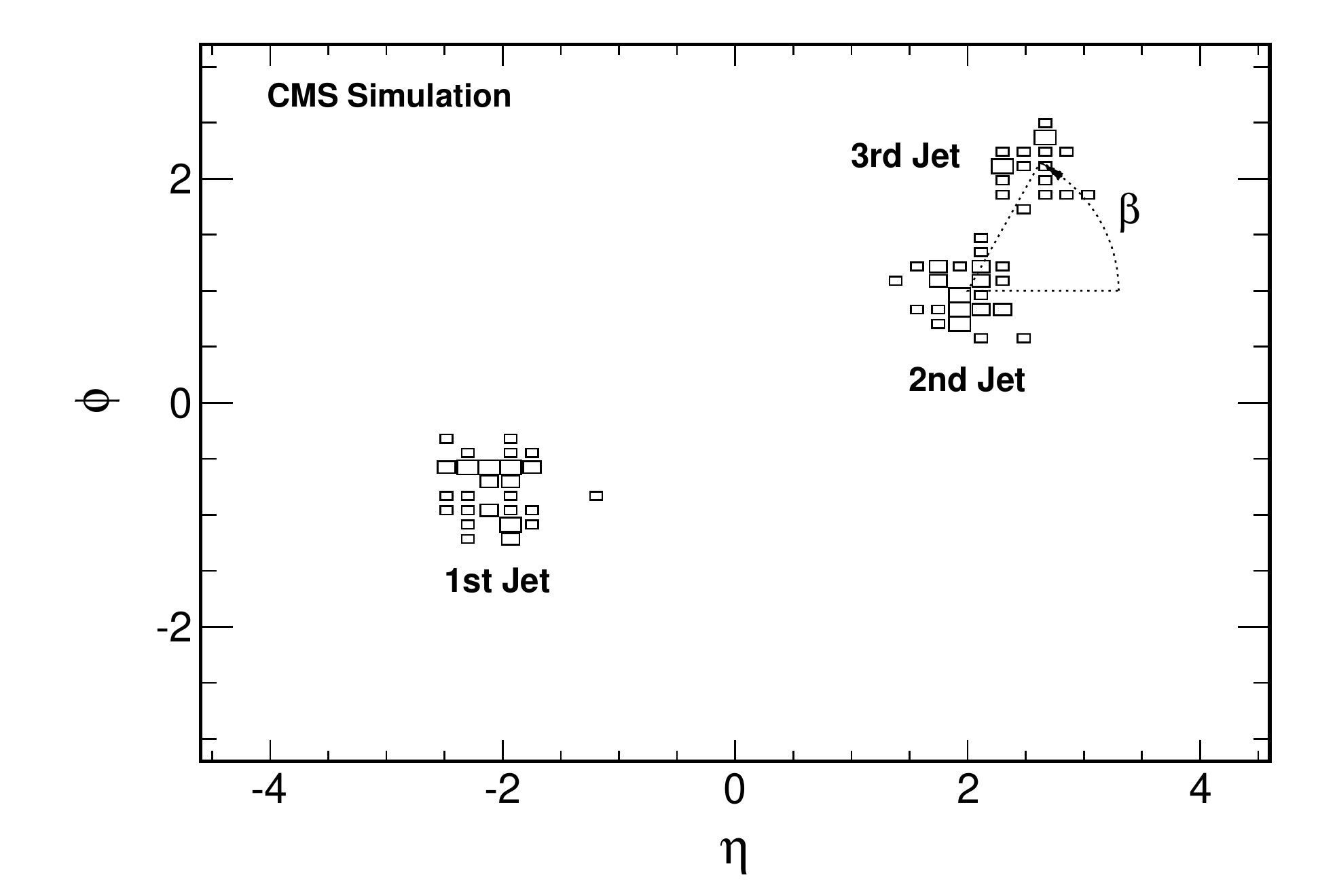}}
\caption{\label{fig:Scheme} Visualization of the observable $\beta$ in ($\eta$,$\phi$) space using a simulated three-jet event. The sizes of the rectangular boxes are proportional to the particle energies.}
\end{center}
\end{figure}

The measured observable $\beta$~\cite{D0} is defined as the azimuthal angle of the third jet with respect to the second jet in ($\eta$, $\phi$) space as shown in Fig.~\ref{fig:Scheme}. Implicitly, this can be expressed by

\begin{equation}
\tan \beta = \frac{\abs{\Delta\phi_{23}}}{\Delta\eta_{23}},
\label{eq1}
\end{equation}

where $\Delta\phi_{23} = \phi_3 - \phi_2$ (defined so that $-\pi \leq \Delta\phi_{23} \leq \pi$), $\Delta\eta_{23} = \sign(\eta_2) \cdot (\eta_3 - \eta_2)$, and $0 \leq \beta \leq \pi$. The absolute value of $\Delta\phi_{23}$ in Eq.~\ref{eq1} and the sign of the pseudorapidity of the second jet, $\sign(\eta_2)$, in the definition of $\Delta\eta_{23}$ are introduced to map symmetric configurations around $\Delta\phi_{23} = 0$ or $\eta = 0$ onto the same $\beta$ value. For $\Delta\phi_{23} = 0$, $\beta$ is defined to be zero or $\pi$ depending on the sign of $\Delta\eta_{23}$ being positive or negative.
In the case of $\Delta\eta_{23} = 0$, which cannot happen simultaneously with $\Delta\phi_{23} = 0$, $\beta$ is defined to equal $\pi/2$.

In a naive leading-order model the two partons are produced back-to-back in the transverse plane.
One of the two partons may radiate a third parton.
In the absence of color coherence effects there is no preferred direction of emission of this third parton around the radiating parton.
In contrast, when color coherence effects are present, the third parton will tend to lie in the event plane defined by the emitting parton and the beam axis.
Therefore, in the presence of color coherence, the third jet population along the event plane (in particular near $\beta \approx 0$) will be enhanced and
out of the plane ($\beta \approx \pi/2$) will be suppressed.
The color coherence effects are expected to become stronger in the region between the second jet and the remnant when the angle between them becomes smaller.
Therefore the study of the $\beta$ variable is performed in two situations: when the second jet is rather central ($\abs{\eta_2} \leq 0.8$) and
when the second jet is more forward ($0.8 < \abs{\eta_2} \leq 2.5$).

The aims of this paper are
\begin{itemize}
\item To measure the $\beta$ distributions, normalized to the total number of events in each region, as a function of $\beta$ separately in the central ($\abs{\eta_2} \leq 0.8$) and forward region ($0.8<\abs{\eta_2} \leq 2.5$):
\begin{equation}
F_{\eta_2, i}(\beta) = \frac{N_{\eta, i}}{N_{\eta}},
\end{equation}
where $N_{\eta}$ is the total number of events in the $\eta_2$ region, $N_{\eta, i}$ the number of events in the given $i$th $\beta$ bin of the $\eta_2$ region.
The choice of this normalization significantly reduces the impact of experimental systematic uncertainties such as the uncertainty in the luminosity.
\item To gauge the sensitivity of the variable $\beta$ to color coherence effects.
\item To compare our measurements to the predictions of MC event generators with various implementations of color coherence.
\end{itemize}

\section{The CMS detector} \label{sec-detector}
A detailed description of the CMS experiment can be found elsewhere~\cite{JINST}; so here we describe the detector systems most relevant to the present analysis.
The central feature of the CMS apparatus is a superconducting solenoid of 6~m internal diameter, providing a magnetic field of 3.8\unit{T}.
Within the field volume, a silicon pixel and strip tracker, a lead tungstate crystal electromagnetic calorimeter
and a brass/scintillator hadron calorimeter (HCAL) are installed.
The central tracking system provides coverage up to $\abs{\eta}= 2.5$ in pseudorapidity and the calorimeters up to $\abs{\eta}=3.0$.
An iron and quartz-fiber Cherenkov forward hadron calorimeter (HF) covers the pseudorapidity range $3.0 < \abs{\eta}< 5.0$.

\section{Event selection}
The CMS detector records events using a two-level trigger system consisting of a hardware-based level-1 (L1) trigger and a software-based high-level trigger (HLT).
For this study, single jet triggers that reconstruct jets from calorimeter energy deposits at L1 and HLT are used to select events based on different $\pt$ jet thresholds.
Five different triggers with $\pt$ thresholds of 30, 50, 70, 100, and 140\GeV are used to select the events.
The triggers were prescaled during the 2010 run when the associated rate exceeded the allocated band width except the highest-threshold one.
Therefore, the events are split into five different bins in $p_\mathrm{T1}$ with each bin containing the events collected during a period when the appropriate trigger was not prescaled.
Each bin starts at $p_{\mathrm{T}\text{min}}$ defined in such a way that the associated trigger efficiency exceeds 99\%.
Table \ref{table:PtInterval} lists the binning in $p_\mathrm{T1}$, and, for each bin, it gives the associated trigger, the number of selected events, and the integrated luminosity for the period during which the given trigger was not prescaled.

Jets are reconstructed with the anti-\kt algorithm~\cite{AK5}, which is implemented in the {\sc FastJet} package~\cite{Cacciari:2011ma} using a distance parameter $R = 0.5$, from a list of particle candidates reconstructed using the particle-flow (PF) algorithm.
This PF algorithm~\cite{PFPAS2009} reconstructs all particle candidates in each event using an optimized combination of information from all CMS subdetector systems:
muons, electrons (with associated bremsstrahlung photons), photons (unconverted and converted), and charged/neutral hadrons. The four-vectors of the neutral
particles are computed by assuming that they come from the primary vertex, which is defined as the vertex with the highest sum of transverse momenta
of all reconstructed tracks pointing to it.
The reconstructed jet energy $E$ is defined as the scalar sum of the energies of the constituents, and the jet momentum $\vec{p}$ is the vector sum of the momenta of the constituents.
The jet transverse momentum \pt is the component of $\vec{p}$ perpendicular to the beam.
The $E$ and $\vec{p}$ values of a reconstructed jet are further corrected for the response of the detector, which is obtained from MC simulations, test beam results, and pp collision data~\cite{CMS_PAS_JME10-010,CMS_PAS_JME10-014}.
The corrections account for the presence of multiple pp collisions in the same or adjacent bunch crossings (pileup interactions) using the jet area method~\cite{Cacciari:2007fd}.

Events are required to have a primary vertex reconstructed within 24\unit{cm} of the detector center along the beam line~\cite{CMS_PAS_TRK10-005}.
Additional selection criteria are applied to each event to remove any spurious jet-like features originating from isolated noise patterns in certain HCAL regions~\cite{CMS_PAS_JME09-008}.
Events having at least three jets with $\pt > 30$ \GeV are selected.
The pseudorapidity of the two leading jets must be within $\abs{\eta_{1}}, \abs{\eta_{2}} \leq 2.5$, while for the third jet no constraints are applied in order to avoid a bias in the $\beta$ measurement.

To further reduce the background from misidentified jets, i.e., jets resulting from noise in the electromagnetic, hadron and/or hadron forward calorimeters, a set of tight identification criteria are applied: each jet should contain at least two particles, one of which is a charged hadron, and the jet energy fraction carried by neutral hadrons, photons, muons, and electrons should be less than 90\%.
With these criteria the contamination of the sample with misidentified jets is suppressed to a level less than 1\%~\cite{PFPAS2009}.

{\tolerance=800
The dijet invariant mass of the two leading jets, $M_{12}$, is required to exceed 220\GeV to ensure a back-to-back configuration.
With this requirement more than 98\% of the events have $\abs{\Delta \phi_{12} - \pi} <1$.
Finally the distance in the ($\eta$, $\phi$) space between the second and third jets is constrained to be $0.5<\Delta R_{23} = \sqrt{(\Delta\eta_{23})^2 + (\Delta\phi_{23})^2}<1.5$ in order to ensure a three-jet topology where the third jet is closer to the second jet.\par}

\begin{table}[htb]
\topcaption{\label{table:PhaseSpace} Summary of the event selection.}
\begin{center}
\begin{tabular}{c} \hline
Selection criteria  \\ \hline
$p_\mathrm{T1}>100$\GeV, $p_\mathrm{T3}>30$\GeV  \\
$\abs{\eta_1}, \abs{\eta_2} \leq 2.5$ \\
$M_{12} > 220$\GeV   \\
$0.5<\Delta R_{23}<1.5$  \\ \hline
\end{tabular}
\end{center}
\end{table}

\begin{figure*}[tbhp]
\begin{center}
{\includegraphics[width=\linewidth]{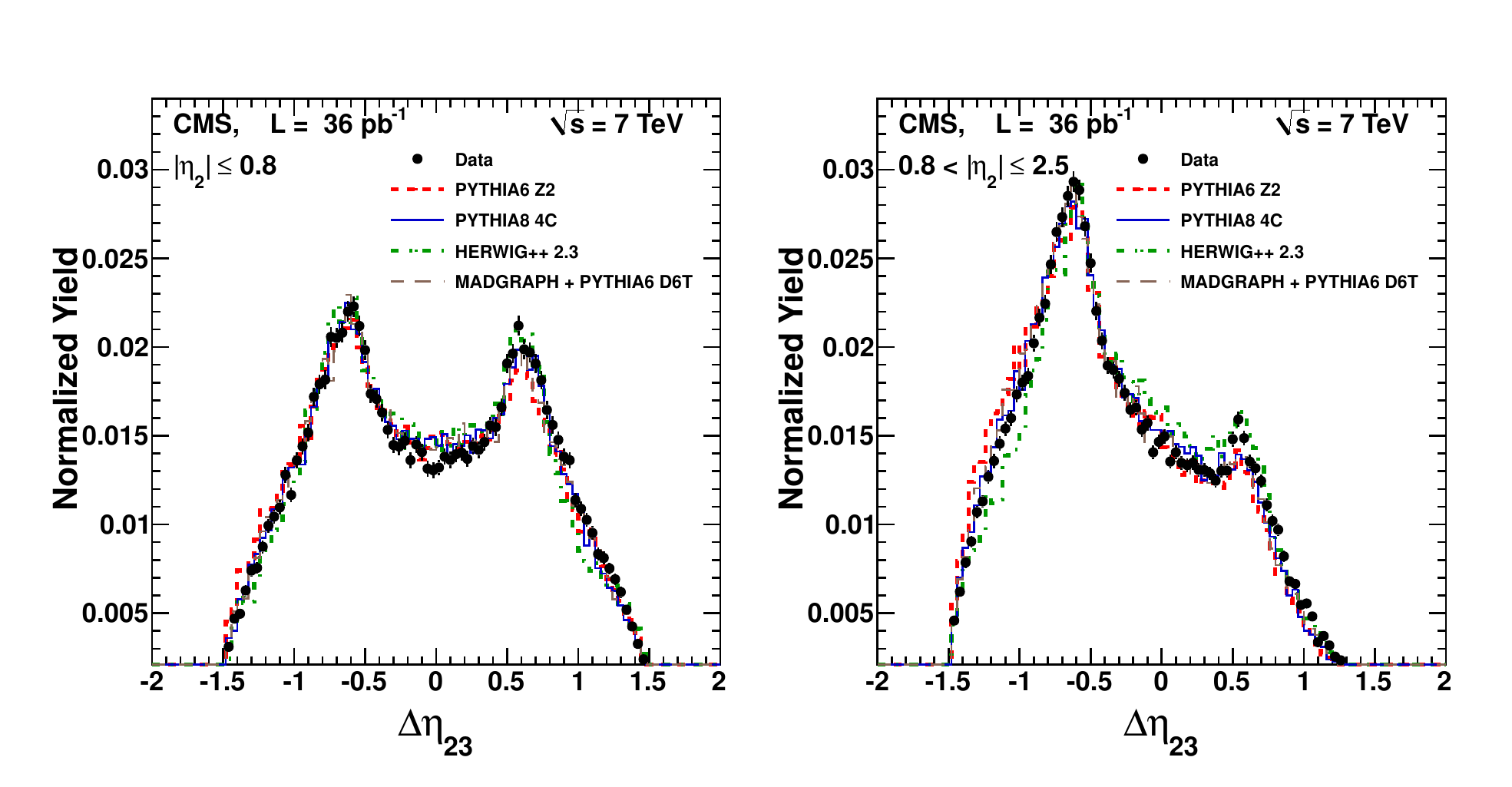}}
\caption{\label{fig:ComDeltaEta} Observed $\Delta \eta_{23}$ distributions, corrected for detector effects, compared to MC predictions by \PYTHIA~6, \PYTHIA~8, \HERWIGpp, and \MADGRAPH + \PYTHIA~6. The MC samples are normalized to the total number of events in data.}
\end{center}
\end{figure*}

\begin{figure*}[tbhp]
\begin{center}
{\includegraphics[width=\linewidth]{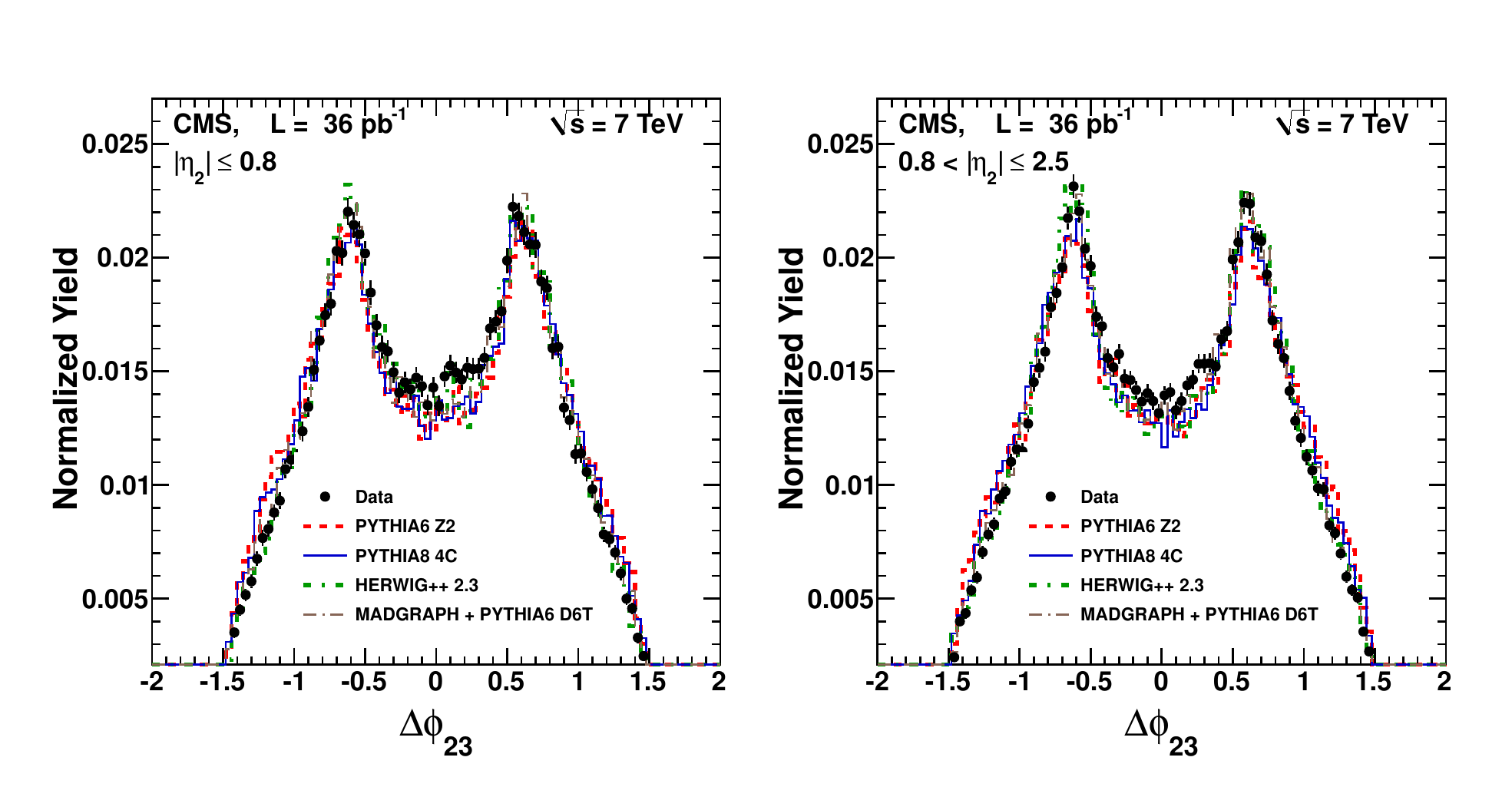}}
\caption{\label{fig:ComDeltaPhi} Observed $\Delta \phi_{23}$ distributions, corrected for detector effects, compared to MC predictions by \PYTHIA~6, \PYTHIA~8, \HERWIGpp, and \MADGRAPH + \PYTHIA~6. The MC samples are normalized to the total number of events in data.}
\end{center}
\end{figure*}

The selections used in the analysis are summarized in Table~\ref{table:PhaseSpace}.
The numbers of events passing the selection criteria in each $p_\mathrm{T1}$ bin are summarized in Table~\ref{table:PtInterval}.
The measured $\Delta \eta_{23}$ and $\Delta \phi_{23}$ distributions are compared to various MC models in Figs.~\ref{fig:ComDeltaEta} and \ref{fig:ComDeltaPhi}.
In general a reasonable agreement is observed with the different models.
A study of the amount of energy collected by the HF detector indicated that there is no diffractive component in the data sample.


\begin{table*}[htb]
\topcaption{\label{table:PtInterval} The binning in $p_\mathrm{T1}$ and, for each bin, the associated trigger, the integrated luminosity for the period during which the
given trigger was not prescaled, and the number of selected events. The selection criteria are described in Table \ref{table:PhaseSpace}.}
\begin{center}
\begin{tabular}{c|ccccc} \hline
$p_\mathrm{T1}$ bin edges & Trigger online & $\lumi_\text{int}$ & \multicolumn{3}{|c}{Number of events} \\
(\GeVns{}) & threshold (\GeVns{})& (\!\pbinv) & \multicolumn{1}{|c}{Total} & $\abs{\eta_2} \leq 0.8$ & $0.8<\abs{\eta_2} \leq 2.5$ \\ \hline

100--120    & 30 &  0.35  &   4511 &1671& 2840 \\
120--160    & 50 & 4.5    &  67 086 &27 069& 40 017 \\
160--200    & 70 & 9.2    &  50 071 &23 055& 27 016 \\
200--250    & 100 &  20     &  39 464 &18 987& 20 477 \\
${>}250$    & 140 & 36     &  31 999 &16 728& 15 271 \\
All	   &	  &      & 193 131 & 87 510& 105 621 \\ \hline
\end{tabular}
\end{center}
\end{table*}

\section{Monte Carlo models}
The reconstructed jets are compared to the predictions of four different Monte Carlo generators that simulate jet production in pp collisions at $\sqrt{s} = 7\TeV$.
The numbers of events for all generator samples is much higher than the number of collected data events so the statistical uncertainties in the MC predictions are not visible in the figures.

The \PYTHIA ~\cite{Pythia6-0} (version 6.422) event generator uses leading-order (LO) matrix elements to generate the 2 $\rightarrow$ 2 hard process in perturbative QCD (pQCD) and the parton shower (PS) model to simulate higher-order processes~\cite{nucl.phys.289_810-846, Phys.Lett.B.185_435-440, epjc_v39_pp129-154}.
The PS model gives a good description of parton emission when the emitted partons are close in phase space.
Events are generated with the Z2 tune for the underlying event.
This Z2 tune is identical to the Z1 tune described in Ref.~\cite{tunesZ2}, except that Z2 uses the CTEQ6L1~\cite{CTEQ6L1} parton distribution functions (PDFs) of the proton in which the parton showers are ordered in $\pt$.
The hadronization is simulated using the Lund string model~\cite{LS-0, LS-1}.
The older D6T tune~\cite{Field:2008zz, Field:2010su, prd_v36_p2019}, where parton showers are ordered in $Q^2$, is considered for comparison.
The D6T tune was designed to describe the lower-energy results of UA5 and CDF\@.
The color coherence effects are implemented in \PYTHIA~6 by means of an angular ordering algorithm where the effects can be switched on and off via the steering parameters MSTP(67) and MSTJ(50), which control the initial-state and the final-state showers, respectively.

The \PYTHIA~8~\cite{Pythia8} (version 8.145) event generator, used with tune 4C~\cite{Tune4C}, orders the parton showers in $\pt$ and models the underlying event using the multiple-parton interaction model from \PYTHIA~6 including initial- and final-state QCD radiation.
The color coherence effects are implemented in a similar manner as for the $\pt$-ordered showers in \PYTHIA~6.

The \HERWIGpp~\cite{PS-0, Herwig++} (version 2.4.2) event generator takes LO matrix elements and simulates parton showers using the coherent branching algorithm with angular ordering of showers.
The cluster hadronization model \cite{Webber1984492} is used in the formation of hadrons from the quarks and gluons produced in the parton shower.
The underlying event is simulated using the eikonal multiple partonic scattering model~\cite{eikonal}.
The color coherence effects are implemented by the angular ordering of emissions in the parton shower using the coherent branching algorithm~\cite{cohe.bran.algo}.

The \MADGRAPH~4~\cite{Madgraph_Alwall:2007} (version 2.24) event generator is interfaced with \PYTHIA~6 for the parton showering and the hadronization using the D6T tune and uses fixed-order matrix element calculations for the multiparton topologies. From two to four partons are considered in the final state.
The color coherence for the hard jets at leading order comes from the exact QCD color amplitudes in the model.
The \kt MLM matching scheme \cite{Mrenna:2003if} applied with a matching parameter of 60\GeV avoids double-counting between the partons from \MADGRAPH and the PS.
\section{Measurement of the normalized \texorpdfstring{$\beta$}{beta} distribution and systematic uncertainties}

The measurement of the $\beta$ distribution is performed in two regions defined by the pseudorapidity of the second jet: the central region $\abs{\eta_2} \leq 0.8$ and the forward region $0.8<\abs{\eta_2} \leq 2.5$.
The angular correlation effects considered in this analysis appear to have a reduced sensitivity to the transverse momentum of the leading jet $p_\mathrm{T1}$.
Consequently different $p_\mathrm{T1}$ bins are merged into one single bin.

The $\beta$ distribution in a given $\eta_2$ region is obtained as a sum of the events weighted by the luminosity collected by the trigger used in the associated $p_\mathrm{T1}$ bin. In case of MC samples the $\beta$ distribution is obtained by summing together the events weighted by their generation level weight in a given $\eta_2$ region.
The normalized $\beta$ distribution is then obtained by dividing the weighted number of events in a given bin of $\beta$ by the total weighted number of events in the given $\eta_2$ region.

In order to correct for the smearing effects induced by the detector resolution, an unfolding procedure is performed using the response matrices obtained from MC event generators.
For this purpose the events generated with the MC programs (\PYTHIA~6, \PYTHIA~8, \MADGRAPH + \PYTHIA~6, and \HERWIGpp) are processed through a full CMS detector simulation package based on \GEANT4~\cite{Geant4}.

Particle-level jets are built from the four-vectors of the MC generated particles with hadronization, but without detector effects.
These jets are obtained using the same jet algorithm as for the reconstructed events.
The resolutions in $\Delta \eta_{23}$ and $\Delta \phi_{23}$ are found to be of the order of 0.005 to 0.01, depending on the transverse momentum and pseudorapidity of the jets.

An iterative Bayesian unfolding technique~\cite{UNFOLD} implemented in the RooUnfold package~\cite{RooUnfold} is used to derive the unfolding corrections to the measured $\beta$ distributions from the detector effects.
The response matrix used to unfold the data is built using \HERWIGpp.
The impact of the unfolding on the normalized distributions is typically of the order of 1\%.

Most of the systematic effects cancel out in the normalized $\beta$ distribution, but the residual influence of several sources of systematic uncertainty has been considered:
\begin{itemize}
\item The jet energy scale uncertainty is evaluated varying the jet response by 2.5--5\%, depending on the $\eta$ and $\pt$ of the jets~\cite{JINST6}.
The impact of this source of systematic uncertainties is below $1\%$.
\item The jet energy and angular resolutions are accounted for by varying them by $\pm 10\%$~\cite{CMS_PAS_JME10-003} and rebuilding
the response matrices for the unfolding accordingly. The observed impact from both sources is in the range of 0.4--0.6\%.
\item The uncertainty due to the unfolding procedure is estimated by the dependence of the response matrix on the choice of MC generator, Alternative response matrices
are built using alternative generators: \PYTHIA~6, \PYTHIA~8 and \MADGRAPH + \PYTHIA~6.
The observed effect is of the order of 0.5\%.
\end{itemize}

The measurement is found to be insensitive to the number of pileup interactions within statistical fluctuations. In the data corresponding
to this analysis the average number of pileup events per bunch crossing was around two.
The total systematic uncertainties for each bin are about 2\%, and a list of the major uncertainties is summarized in Table~\ref{table:Systematic}.
Each systematic source was found to be fully correlated between $\beta$ and $\eta_2$ bins~\cite{JINST6, CMS_PAS_JME10-003}.
However, the various systematic sources are uncorrelated among themselves.

\begin{table*}[htbp]
\topcaption{\label{table:Systematic} Typical systematic and statistical uncertainties in the normalized $\beta$ spectrum and the statistical errors.}
\begin{center}
\begin{tabular}{l | c  c} \hline
Uncertainty sources  & $\abs{\eta_2} \leq 0.8$  & $0.8<\abs{\eta_2} \leq 2.5$ \\ \hline
Jet energy scale (JES)&    1.0\%       &   1.0\%       \\
Jet energy resolution (JER)&    0.4\%       &   0.5\%       \\
Jet angular resolution (JAR)&    0.5\%       &   0.6\%       \\
Physics model (PM) used in unfolding &    0.6\%       &   0.7\%       \\ \hline
Statistical uncertainty  &    4.0\%       &   3.7\%       \\ \hline
\end{tabular}
\end{center}
\end{table*}

\begin{table*}[htbp]
\topcaption{\label{table:FinalResultCen} The unfolded $\beta$ distributions and their uncertainties for the central region $\abs{\eta_2} \leq 0.8$. All uncertainties are symmetric and given in percent ($\%$).}
\begin{center}
\begin{tabular}{c|ccccccc} \hline
$\beta$ (degree) & $F_{\eta_2}(\beta)$ & $\sigma_\mathrm{Stat}$ & $\sigma_\mathrm{JES}$ & $\sigma_\mathrm{JER}$ & $\sigma_\mathrm{JAR}$ & $\sigma_\mathrm{PM}$ & $\sigma_\text{Syst}$\\ \hline
0--10 &  0.0549 & 3.5 & 1.0    & 0.3   & 0.4 & 0.6 & 1.3\\
10--20 &  0.0535 & 3.9 & 1.1    & 0.4   & 0.6 & 0.6 & 1.4 \\
20--30 &  0.0544 & 4.2 & 0.5    & 0.5   & 0.3 & 0.6 & 1.0\\
30--40 &  0.0538 & 4.0 & 1.1    & 0.2   & 0.3 & 0.6 & 1.3\\
40--50 &  0.0525 & 3.8 & 0.5    & 0.5   & 0.5 & 0.6 & 1.1\\
50--60 &  0.0515 & 4.4 & 0.6    & 0.6   & 0.7 & 0.6 & 1.3\\
60--70 &  0.0515 & 4.3 & 0.6    & 0.4   & 0.6 & 0.6 & 1.1\\
70--80 &  0.0519 & 4.1 & 0.5    & 0.3   & 0.4 & 0.6 & 0.9\\
80--90 &  0.0511 & 4.2 & 0.4    & 0.4   & 0.5 & 0.6 & 1.0\\
90--100 & 0.0515 & 4.3 & 0.5    & 0.3   & 0.2 &  0.6 & 0.9\\
100--110 & 0.0528 & 4.3 & 0.5    & 0.4   & 0.5 &  0.6 & 1.0\\
110--120 & 0.0543 & 4.3 & 0.6    & 0.6   & 0.3 &  0.6 & 1.1\\
120--130 & 0.0580 & 4.1 & 1.2    & 0.5   & 0.4 &  0.6 & 1.5\\
130--140 & 0.0583 & 3.7 & 0.5    & 0.6   & 0.3 &  0.6 & 1.0\\
140--150 & 0.0616 & 4.2 & 0.6    & 0.5   & 0.5 &  0.6 & 1.1\\
150--160 & 0.0622 & 3.9 & 0.9    & 0.6   & 0.5 &  0.6 & 1.3\\
160--170 & 0.0626 & 3.6 & 0.7    & 0.5   & 0.6 &  0.6 & 1.2\\
170--180 & 0.0638 & 3.2 & 0.5    & 0.7   & 0.6 &  0.6 & 1.2\\ \hline
\end{tabular}
\end{center}
\end{table*}

\begin{table*}[htbp]
\topcaption{\label{table:FinalResultFor} The unfolded $\beta$ distributions and their uncertainties for the forward region $0.8<\abs{\eta_2} \leq 2.5$. All uncertainties are symmetric and given in percent ($\%$).}
\begin{center}
\begin{tabular}{c|ccccccc} \hline
$\beta$ (degree) & $F_{\eta_2}(\beta)$ & $\sigma_\mathrm{Stat}$ & $\sigma_\mathrm{JES}$ & $\sigma_\mathrm{JER}$ & $\sigma_\mathrm{JAR}$ & $\sigma_\mathrm{PM}$ & $\sigma_\text{Syst}$\\ \hline
0--10 &  0.0388 & 3.9 & 1.6    & 0.5   & 0.5 & 0.7 & 1.9\\
10--20 &  0.0391 & 4.6 & 0.6    & 0.5   & 0.6 & 0.7 & 1.2 \\
20--30 &  0.0406 & 4.4 & 0.7    & 0.4   & 0.5 & 0.7 & 1.2\\
30--40 &  0.0404 & 4.6 & 0.5    & 0.4   & 0.5 & 0.7 & 1.1\\
40--50 &  0.0414 & 4.2 & 0.6    & 0.5   & 0.5 & 0.7 & 1.2\\
50--60 &  0.0438 & 3.9 & 0.7    & 0.4   & 0.4 & 0.7 & 1.1\\
60--70 &  0.0430 & 4.4 & 0.8    & 0.5   & 0.6 & 0.7 & 1.3\\
70--80 &  0.0476 & 4.2 & 0.5    & 0.5   & 0.6 & 0.7 & 1.2\\
80--90 &  0.0491 & 4.0 & 1.2    & 0.4   & 0.5 & 0.7 & 1.5\\
90--100 & 0.0520 & 3.9 & 0.8    & 0.5   & 0.4 &  0.7 & 1.2\\
100--110 & 0.0567 & 3.6 & 0.8    & 0.5   & 0.5 &  0.7 & 1.3\\
110--120 & 0.0625 & 3.5 & 0.7    & 0.5   & 0.5 &  0.7 & 1.2\\
120--130 & 0.0662 & 3.2 & 0.8    & 0.5   & 0.6 &  0.7 & 1.3\\
130--140 & 0.0692 & 3.2 & 0.7    & 0.4   & 0.6 &  0.7 & 1.2\\
140--150 & 0.0736 & 3.1 & 0.6    & 0.6   & 0.5 &  0.7 & 1.2\\
150--160 & 0.0774 & 2.9 & 0.7    & 0.4   & 0.6 &  0.7 & 1.2\\
160--170 & 0.0795 & 2.9 & 0.8    & 0.5   & 0.5 &  0.7 & 1.3\\
170--180 & 0.0791 & 2.6 & 0.8    & 0.6   & 0.5 &  0.7 & 1.3\\ \hline
\end{tabular}
\end{center}
\end{table*}

\section{Results}
\begin{figure*}[tbhp]
\begin{center}
{\includegraphics[width=0.95\linewidth]{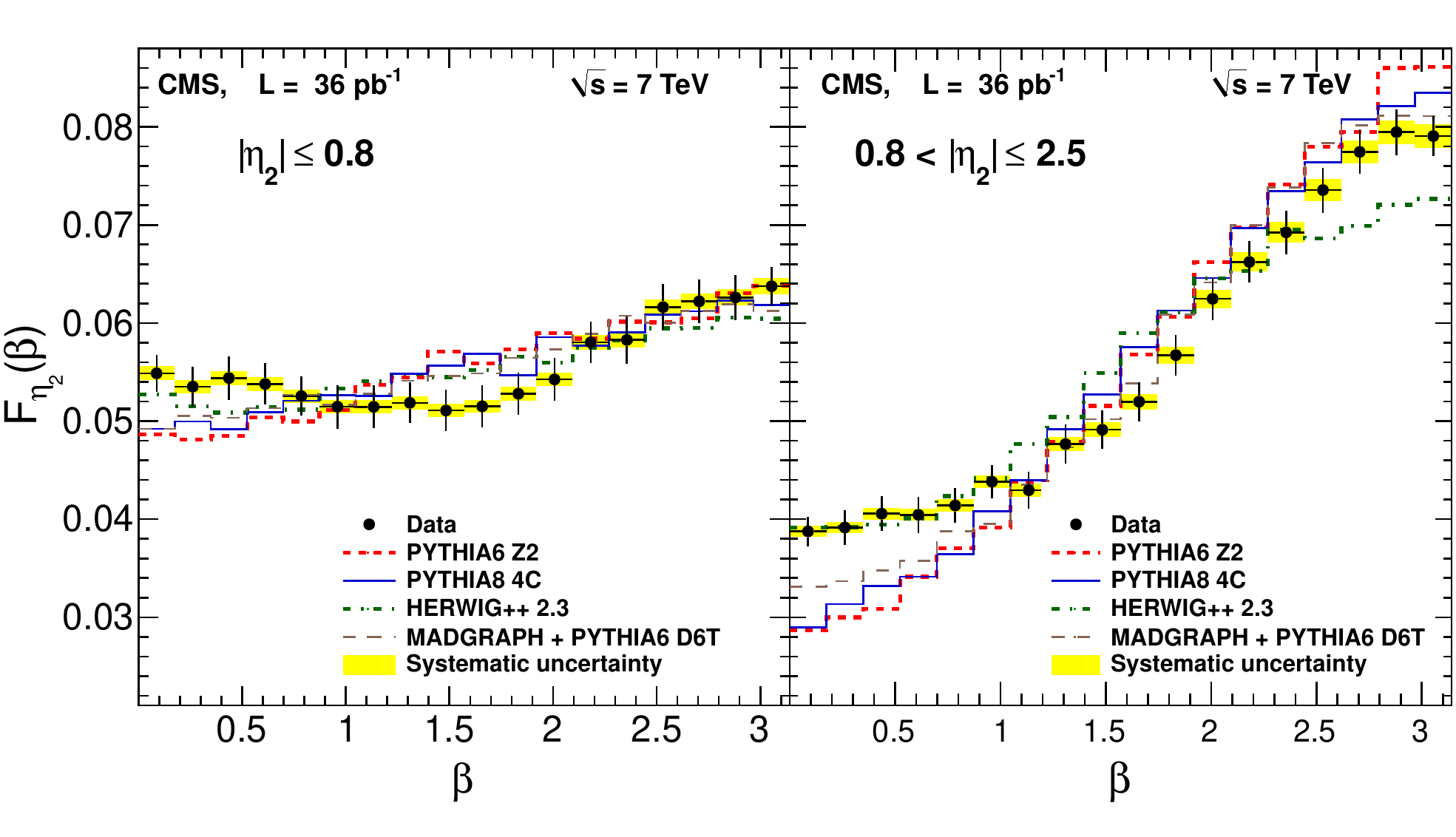}}
\caption{\label{fig:UnfoldDataOverMC} Observed $\beta$ distributions for the data, corrected for detector effects, and for the MC generators (\PYTHIA~6, \PYTHIA~8, \HERWIGpp, and \MADGRAPH + \PYTHIA~6) in the central ($\abs{\eta_{2}} \leq 0.8$) and forward ($0.8 < \abs{\eta_2} \leq 2.5$) regions. The error bars show the statistical uncertainties, while the yellow shaded bands correspond to the combined systematic uncertainty.}
\end{center}
\end{figure*}

The unfolded $\beta$ distributions are shown in Fig.~\ref{fig:UnfoldDataOverMC} together with the predictions from the various MC models for the central ($\abs{\eta_2} \leq 0.8$) and forward ($0.8 < \abs{\eta_2} \leq 2.5$) regions.
The values of the unfolded $\beta$ distributions and their uncertainties are presented in Tables~\ref{table:FinalResultCen} and \ref{table:FinalResultFor}.

\begin{figure*}[tbhp]
\begin{center}
{\includegraphics[width=0.9\linewidth]{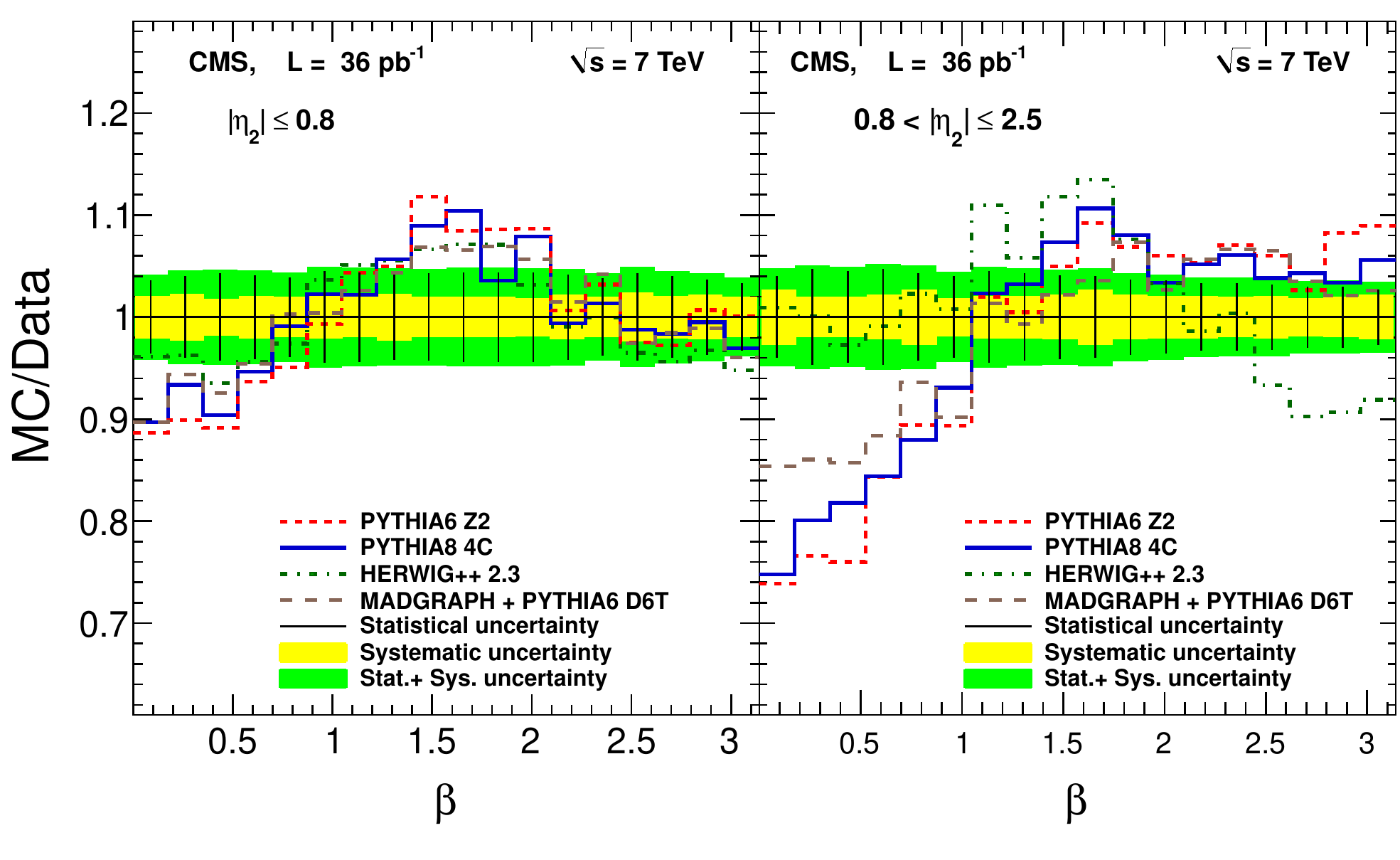}}
\caption{\label{fig:FinalBeta} The ratio of the various MC predictions to the measured $\beta$ distribution. The error bars show the statistical uncertainty of the data. The yellow band represents the systematic uncertainty, while the green band represents the total uncertainty. }
\end{center}
\end{figure*}

The ratios of the various MC predictions to the measured $\beta$ distributions are shown in Fig.~\ref{fig:FinalBeta}.
The data exhibit a clear enhancement of events compared to the \PYTHIA and \MADGRAPH generators near the event plane ($\beta = 0$) and a suppression in the transverse plane ($\beta = \pi/2$).
The $\chi^2$ comparisons of data with MC simulation, taking into account the statistical and systematic correlations between different data points, are shown separately for the central and forward regions in Table~\ref{table:Chisq}.
The number of degrees of freedom (NDF) is 17, which is the number of bins minus one to account for the constraint imposed by the normalization.

None of the models used in the analysis describes the data satisfactorily.
Even though \PYTHIA~6 was adjusted with the Tevatron data, it fails to describe the LHC data since the $\chi^2$/NDF is large.
No significant difference is observed between the tunes D6T and Z2.
The \PYTHIA~8 tune 4C generator describes the data better than \PYTHIA~6 over the entire phase space, but the disagreement in the forward region is not negligible.
The \HERWIGpp event generator describes the data better than the other MC generators in the central region, but the agreement is poor in the forward region.
Finally, when \MADGRAPH is used with the exact $2 \rightarrow 3$ matrix element calculations at LO, the global description of the data is improved with respect to \PYTHIA~6 alone.

\begin{figure*}[htbp]
\begin{center}
{\includegraphics[width=0.95\linewidth]{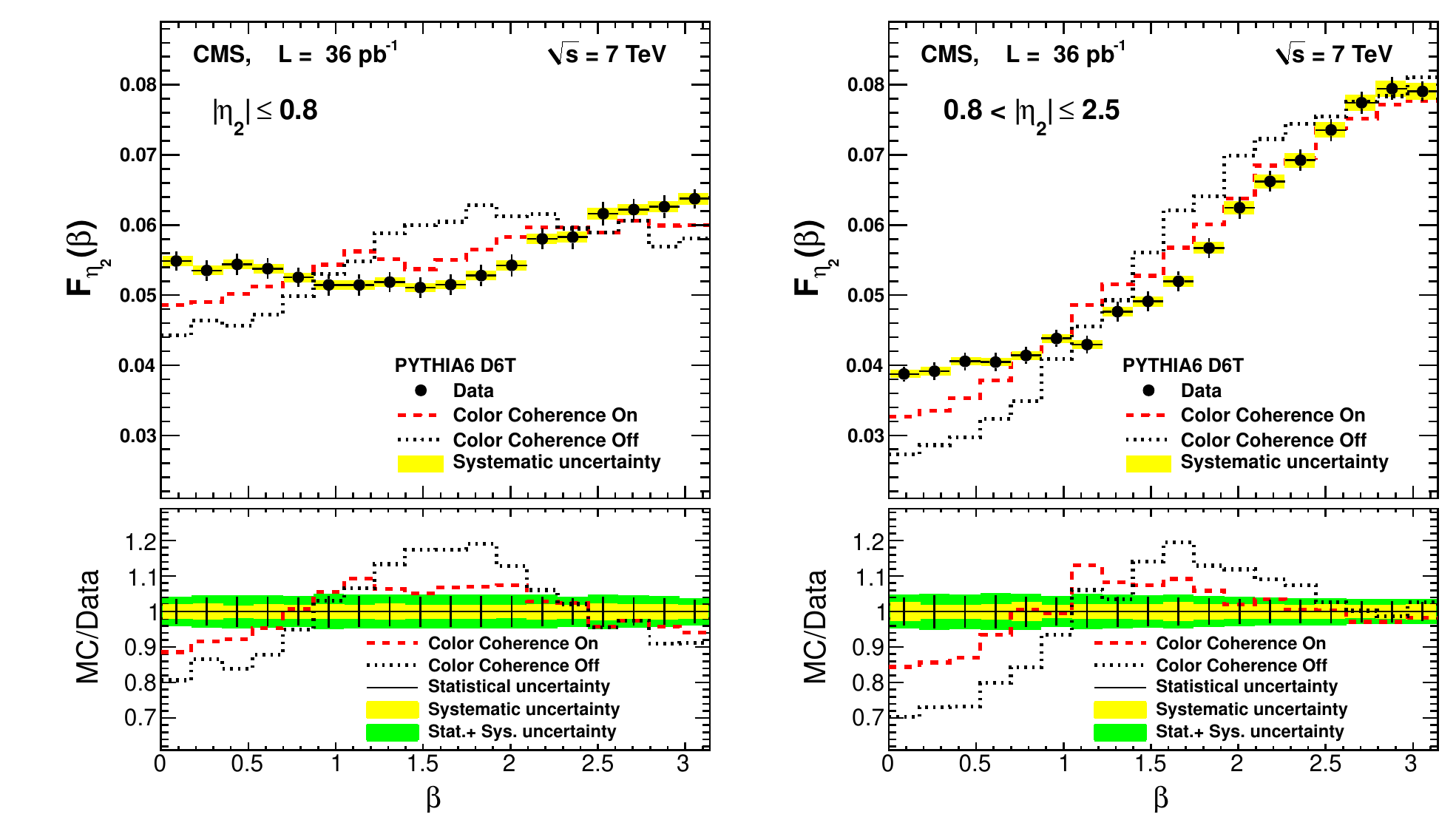}}
\caption{\label{fig:PythiaD6TCC} The MC predictions for the $\beta$ distribution from \PYTHIA~6, with and without color coherence effects in the first branching of the initial- and final-state showers, compared to the measurement. The error bars show the uncorrelated statistical uncertainty of the data. The yellow band represents the systematic uncertainty, while the green band represents the total uncertainty.}
\end{center}
\end{figure*}

The impact of the color coherence effects is studied by switching them on and off for the first emission in the initial- and final-state showers in \PYTHIA~6.
One can observe in Fig.~\ref{fig:PythiaD6TCC} that the agreement between the data and the simulation deteriorates when the color coherence effects in the MC events are suppressed.
More quantitatively, the $\chi^2$ divided by the number of degrees of freedom increases up to 7.7 in the central region and 11.5 in the forward region.
The first emission in the initial- and final-state showers contributes roughly the same order.
Using \PYTHIA, it has been verified that the impact of the non-perturbative component of the QCD calculation (hadronization and underlying event) is negligible for this analysis.
One conclusion from this \PYTHIA study, as shown Fig.~\ref{fig:PythiaD6TCC}, is that the data clearly support larger color coherence effects than in present MC implementations.
\begin{table*}
\topcaption{\label{table:Chisq} Values of $\chi^2$ for comparisons of the $\beta$ distribution for the data with the predictions of various MC generators.
The number of degrees of freedom for both regions is 17. }
\begin{center}
\begin{tabular}{c | c  c} \hline
                                & \multicolumn{2} {c} {$\chi^2$/NDF} \\
MC event generator  & $\abs{\eta_2} \leq 0.8$    &  $0.8<\abs{\eta_2} \leq 2.5$ \\ \hline
\PYTHIA~6 Z2                             &   2.5 &  8.1      \\
\PYTHIA~8 4C                             &  1.7 &  6.4       \\
\HERWIGpp 2.3                        &  1.2  &  3.5       \\
\MADGRAPH + \PYTHIA~6 &  1.6   &  3.3      \\ \hline
\end{tabular}
\end{center}
\end{table*}

\section{Summary}
Color coherence effects in multijet events have been studied in a sample of pp collisions corresponding to an integrated luminosity of 36\pbinv, collected with the CMS detector at $\sqrt{s} = 7\TeV$.
Distributions of the variable $\beta$, which was previously used in similar analyses at the Tevatron, are used to measure the angular correlation between the second and third jets in transverse-momentum order, in the pseudorapidity and azimuthal angle space.
The measurements, unfolded for detector effects, are compared to the predictions of the MC event generators \PYTHIA~6, \PYTHIA~8, \HERWIGpp, and \MADGRAPH + \PYTHIA~6 in the central and forward rapidity regions.
We have shown that the variable $\beta$ is sensitive to color coherence effects, and insensitive to the hadronization and underlying event.
It is necessary to implement the color coherence effects in MC simulations to better describe the data.
Although the MC models in the analysis include this effect by default, none of them describes the data satisfactorily for all $\beta$ values.
The \PYTHIA~6 expectations predict weaker color coherence effects than those observed, while \PYTHIA~8 exhibits a better agreement with the data.
The \MADGRAPH MC generator, which uses the exact $2 \rightarrow 3$ matrix element calculations
at LO matched to \PYTHIA~6 for parton showering, improves the agreement with data with respect to \PYTHIA~6 alone, while \HERWIGpp describes the data in the central region better than the other MC generators but shows discrepancies in the forward region.

\section*{Acknowledgements}
{\tolerance=1200
We congratulate our colleagues in the CERN accelerator departments for the excellent performance of the LHC and thank the technical and administrative staffs at CERN and at other CMS institutes for their contributions to the success of the CMS effort. In addition, we gratefully acknowledge the computing centers and personnel of the Worldwide LHC Computing Grid for delivering so effectively the computing infrastructure essential to our analyses. Finally, we acknowledge the enduring support for the construction and operation of the LHC and the CMS detector provided by the following funding agencies: BMWF and FWF (Austria); FNRS and FWO (Belgium); CNPq, CAPES, FAPERJ, and FAPESP (Brazil); MES (Bulgaria); CERN; CAS, MoST, and NSFC (China); COLCIENCIAS (Colombia); MSES (Croatia); RPF (Cyprus); MoER, SF0690030s09 and ERDF (Estonia); Academy of Finland, MEC, and HIP (Finland); CEA and CNRS/IN2P3 (France); BMBF, DFG, and HGF (Germany); GSRT (Greece); OTKA and NIH (Hungary); DAE and DST (India); IPM (Iran); SFI (Ireland); INFN (Italy); NRF and WCU (Republic of Korea); LAS (Lithuania); CINVESTAV, CONACYT, SEP, and UASLP-FAI (Mexico); MBIE (New Zealand); PAEC (Pakistan); MSHE and NSC (Poland); FCT (Portugal); JINR (Dubna); MON, RosAtom, RAS and RFBR (Russia); MESTD (Serbia); SEIDI and CPAN (Spain); Swiss Funding Agencies (Switzerland); NSC (Taipei); ThEPCenter, IPST, STAR and NSTDA (Thailand); TUBITAK and TAEK (Turkey); NASU (Ukraine); STFC (United Kingdom); DOE and NSF (USA).
\par
}

\bibliography{auto_generated}   

\cleardoublepage \appendix\section{The CMS Collaboration \label{app:collab}}\begin{sloppypar}\hyphenpenalty=5000\widowpenalty=500\clubpenalty=5000\textbf{Yerevan Physics Institute,  Yerevan,  Armenia}\\*[0pt]
S.~Chatrchyan, V.~Khachatryan, A.M.~Sirunyan, A.~Tumasyan
\vskip\cmsinstskip
\textbf{Institut f\"{u}r Hochenergiephysik der OeAW,  Wien,  Austria}\\*[0pt]
W.~Adam, T.~Bergauer, M.~Dragicevic, J.~Er\"{o}, C.~Fabjan\cmsAuthorMark{1}, M.~Friedl, R.~Fr\"{u}hwirth\cmsAuthorMark{1}, V.M.~Ghete, N.~H\"{o}rmann, J.~Hrubec, M.~Jeitler\cmsAuthorMark{1}, W.~Kiesenhofer, V.~Kn\"{u}nz, M.~Krammer\cmsAuthorMark{1}, I.~Kr\"{a}tschmer, D.~Liko, I.~Mikulec, D.~Rabady\cmsAuthorMark{2}, B.~Rahbaran, C.~Rohringer, H.~Rohringer, R.~Sch\"{o}fbeck, J.~Strauss, A.~Taurok, W.~Treberer-Treberspurg, W.~Waltenberger, C.-E.~Wulz\cmsAuthorMark{1}
% ...
\vskip\cmsinstskip
% ...
1:~~Also at Vienna University of Technology, Vienna, Austria\\
2:~~Also at CERN, European Organization for Nuclear Research, Geneva, Switzerland\\
% ...

\end{sloppypar}
\end{document}